\documentclass[reprint,aps,prd,twocolumn,superscriptaddress,nofootinbib,amsmath,amssymb]{revtex4-2}

\usepackage{graphicx}
\usepackage{dcolumn}
\usepackage{bm}
\usepackage{natbib}

\usepackage{color}
\usepackage{hyperref}
\hypersetup{colorlinks=true,linkcolor=blue,filecolor=blue,urlcolor=blue,citecolor=blue}
\usepackage{hyperref}

\newcommand{{\AGN}}{AGN J124942.3 + 344929}
\newcommand{\Msun}{M_{\odot}}

\begin{document}%

\title{Multimessenger hierarchical triple merger gravitational-wave event pair\\ GW190514-GW190521 inside AGN J124942.3 + 344929}

\author{Guo-Peng Li}
\affiliation{
Department of Astronomy, School of Physics and Technology, Wuhan University, Wuhan 430072, China}

\author{Xi-Long Fan}
\email[Corresponding author: Xi-Long Fan\\]{xilong.fan@whu.edu.cn}
\affiliation{
Department of Astronomy, School of Physics and Technology, Wuhan University, Wuhan 430072, China}

\date{\today}

\begin{abstract}

There is a candidate electromagnetic (EM) counterpart to the binary black hole merger GW190521, identified as ZTF19abanrhr within active galactic nuclei (AGN) J124942.3 + 344929. Additionally, GW190514 is proposed as a plausible precursor merger to GW190521 within a hierarchical merger scenario. In this study, we investigate the potential association between GW190514 and GW190521 as a hierarchical triple merger associated with ZTF19abanrhr, taking into account of the sky position, distance, and mass of the sources using a Bayesian criterion. Our analysis reveals that the association is favored over a random coincidence, with a log Bayes factor of 16.8, corresponding to an odds ratio of $\sim$$199:1$, assuming an astrophysical prior odds of $10^{-5}$. Notably, when accounting for the primary masses of the two gravitational-wave events as potential products of mergers in the AGN formation channel, the Bayes factor increases significantly, further enhancing the preference for this association by a factor of $\sim$$10^2$, corresponding to a log Bayes factor of 21.5 and an odds ratio of $\sim$$2\times10^4:1$. Our results suggest strong evidence for the first hierarchical triple merger associated with an EM counterpart in the AGN formation channel. This work is crucial for understanding the formation mechanisms of massive black holes, the role of AGNs in hierarchical mergers, and the implications of multimessenger astronomy.

\end{abstract}

\maketitle

\section{Introduction}
\begin{figure}
\centering
\includegraphics[width=8.5cm]{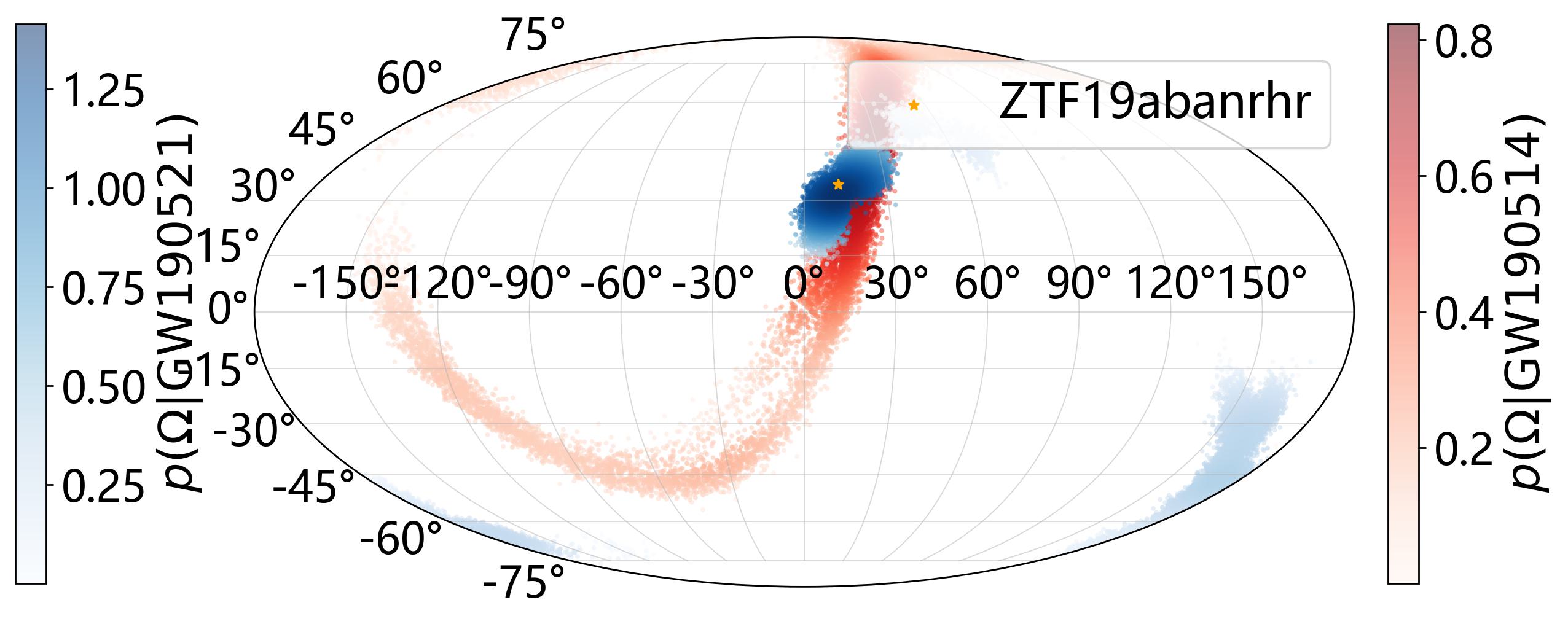}
\includegraphics[width=7cm]{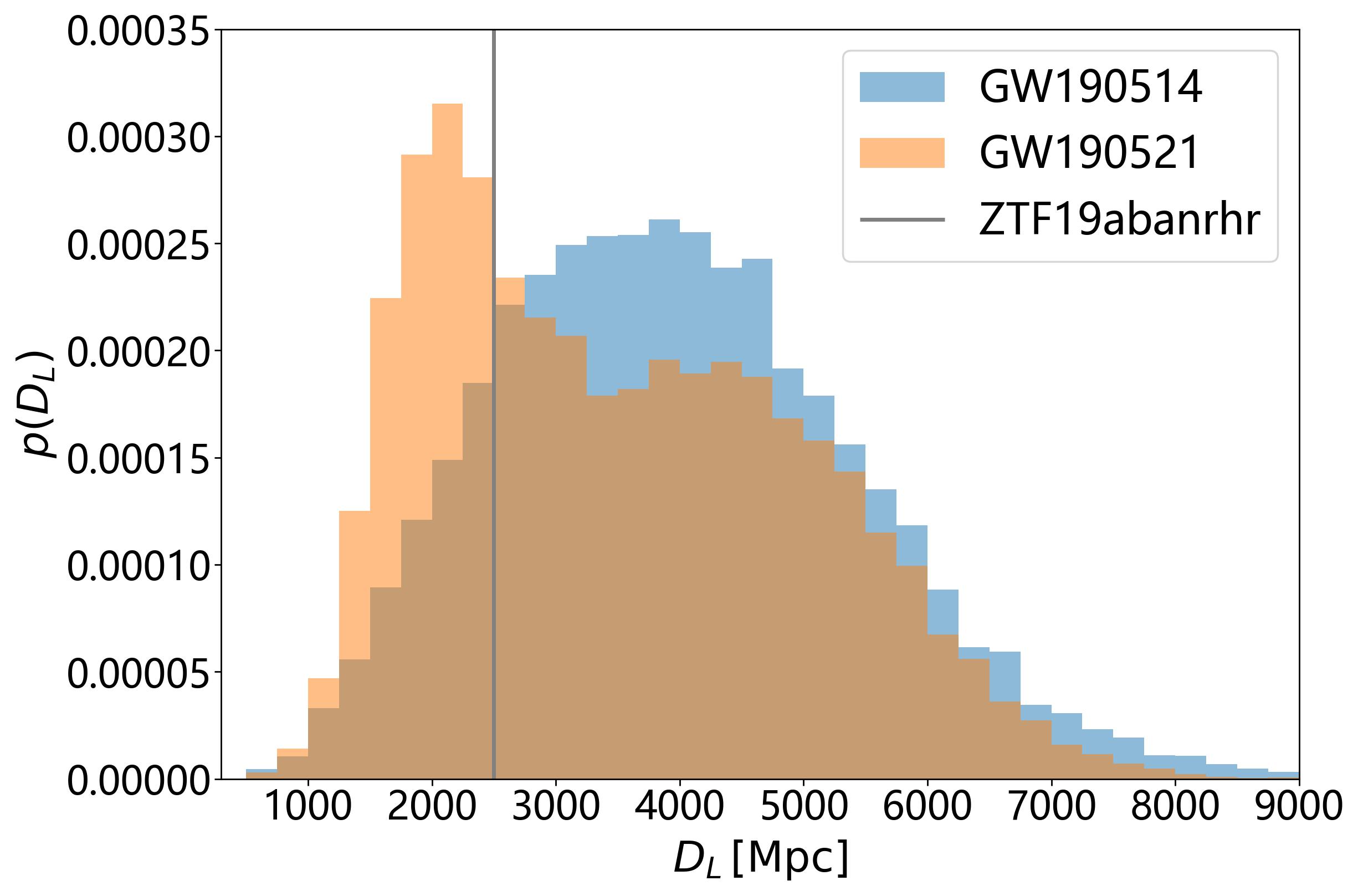}
\caption{
Probability distribution of the sky position and distance.
Top: Mollweide projection of the sky positions of GW190514 (red), GW190521 (blue), and ZTF19abanrhr (orange) within {\AGN}. ZTF19abanrhr lies within the 78\% spatial contour for GW190521's sky localization, while GW190514's 90\% sky localization is consistent with the position of ZTF19abanrhr.
Bottom: Marginalized luminosity distance distributions integrated over the sky for GW190514 (blue) and GW190521 (yellow). The luminosity distance (gray) of ZTF19abanrhr is derived from the redshift $z = 0.438$ of {\AGN}, assuming a Planck18 cosmology~\citep{2020A&A...641A...6P}. 
}
\label{fig1} 
\end{figure}

GW190521~\citep{2020PhRvL.125j1102A} is one of the most massive binary black hole (BBH) mergers observed during the first three LIGO-Virgo observing runs, with component masses of approximately $85\,\Msun$ and $66\,\Msun$, producing a remnant black hole (BH) of around $142\,\Msun$, which can be classified as an intermediate mass BH. GW190514~\citep{2024PhRvD.109b2001A} is another BBH merger, with component masses of approximately $39\,\Msun$ and $28\,\Msun$, resulting in a remnant BH of around $67\,\Msun$, which overlaps with the second mass range of GW190521. Moreover, there is significant spatial overlap between the sky localizations and distances of GW190514 and GW190521~(see~Fig.\,\ref{fig1}), suggesting a plausible scenario where these two events are part of a hierarchical triple merger. In this scenario, GW190514 would represent the first merger, producing a remnant BH that later paired with another BH to form the hierarchical system responsible for GW190521~\citep{2021ApJ...907L..48V}.

In addition, an optical electromagnetic (EM) counterpart candidate, ZTF19abanrhr, was reported by~\citet{2020PhRvL.124y1102G}, detected by the Zwicky Transient Facility approximately 34 days after the gravitational-wave (GW) event GW190521. Evidence in favor of an association between the GW event and the EM counterpart was also reported in \citep{2023PhRvD.108l3039M,2021arXiv211212481C}~(although see \citep{2021ApJ...914L..34P,2021CQGra..38w5004A}).
This counterpart, confirmed as a flare within the active galactic nuclei (AGN) {\AGN}, was at the 78\% spatial contour for GW190521's sky localization. Interestingly, GW190514's 90\% sky localization also coincides with the position of ZTF19abanrhr~(see~Fig.\,\ref{fig1}). This raises an intriguing question: Could both GW190514 and GW190521 be a hierarchical triple merger originating from {\AGN}, with ZTF19abanrhr serving as their EM counterpart?

Theoretical models suggest that AGN disks are promising environments for massive and hierarchical BBH mergers~\citep{2019PhRvL.123r1101Y,2022PhRvD.105f3006L,2023PhRvD.107f3007L,2025ApJ...981..177L}, consistent with the observed events GW190514 and GW190521. In such environments, EM emissions are expected, likely produced by shocks from accreting merger remnants interacting with the baryon-rich, high-density surroundings~\citep{2019ApJ...884L..50M,2021ApJ...916L..17W,2023ApJ...950...13T,2024ApJ...961..206C}. 
Additionally, AGN disks could facilitate an excess of eccentric mergers, which may provide a possible explanation for the non-zero eccentricity and significant spin–orbit tilt of GW190521 if it originated in such an environment~\citep{2022Natur.603..237S}. Moreover, the final mass of GW190514 is consistent with the best-matching eccentric template for GW190521's mass distribution~\citep{2021ApJ...907L..48V}.

Motivated by this, we propose a compelling scenario: GW190514, detected by the Advanced LIGO~\citep{2015CQGra..32g4001L}, represents a first generation BH merger occurring within {\AGN}, which left behind a remnant BH that rapidly merged with another BH, facilitated by interactions with the gas, leading to the formation of GW190521, detected by the Advanced LIGO and the Advanced Virgo~\citep{2015CQGra..32b4001A}. Concurrently, the accreting remnant BH from either GW190521 or GW190514 produced an optical flare ZTF19abanrhr, which was observed by the Zwicky Transient Facility~\citep{Bellm_2019} approximately 40 days after the merger~\citep{2020PhRvL.124y1102G,2023ApJ...942...99G}.

The rest of this paper is organized as follows. In Section~\ref{sec:method}, we provide a description of the methodology used in this study. In Section~\ref{sec:result}, we present the results of our analysis. We conclude with our discussion in Section~\ref{sec:cd}.

\section{Methods}\label{sec:method}
We explore the potential association between GW190514 and GW190521 as a hierarchical triple merger in {\AGN}, associated with ZTF19abanrhr, using data ($d_1$ and $d_2$ for the gravitational signals\footnote{\url{https://zenodo.org/record/5117702}}, and $s$ for the EM signal\footnote{\url{https://lasair-ztf.lsst.ac.uk/objects/ZTF19abanrhr/}}) from the GWTC-2.1 release~\citep{2024PhRvD.109b2001A} and ZTF~\citep{Bellm_2019}. This is a hypothesis-testing question. Our analysis employs a Bayesian criterion to assess whether signals observed from these separate datasets share a common origin. The association ($\mathcal{H}_1$) at least requires that i) the localizations of GW190514, GW190521, and ZTF19abanrhr must be consistent (see~Fig.\,\ref{fig1}); and ii) the final mass of GW190514 should align with the second mass in GW190521. In contrary, the coincidence ($\mathcal{H}_0$) corresponds to the scenario where the three detections are associated by chance. As a result, the \textit{odds} $\mathcal{O}^1_0$ between the hypotheses $\mathcal{H}_1$ and $\mathcal{H}_0$ is given by: $\mathcal{O}^1_0(d_1,d_2,s) = \mathcal{P}^1_0\,\mathcal{B}^1_0(d_1,d_2,s)$, where $\mathcal{P}^1_0$ is the \textit{prior odds}; and $\mathcal{B}^1_0(d_1,d_2,s) = 
p(d_1,d_2,s|\mathcal{H}_1)/p(d_1,d_2,s|\mathcal{H}_0)$
is the \textit{Bayes factor} defined by the ratio of the
evidences of the hypotheses.

The derivations and computational framework presented as follows build upon the foundational methodology established in~\citet{2023PhRvD.108l3039M}, who demonstrated the astrophysical association between GW190521 and flare ZTF19abanrhr as its EM counterpart. Here, we investigate the association between GW190514 and GW190521 as a hierarchical triple merger associated with ZTF19abanrhr.

\subsection{Environmental effects}\label{ee}
Stellar-mass BBHs residing in AGN disks near the central supermassive BH are influenced by various environmental effects. These effects can introduce biases in the parameter estimates of GW events, particularly altering the observed mass and distance of sources. Following the analysis in the previous literature~\citep{2023PhRvD.108l3039M}, both effects impact the properties of the sources.

The first effect, related to the gravitational potential of the supermassive BH, leads to gravitational redshift:
\begin{equation}\label{grav}
z_{\rm grav} = 
\left(1-\frac{R_{\rm s}}{r}\right)^{1/2} - 1\,,
\end{equation}
where $R_{\rm s}$ represents the Schwarzschild radius of the supermassive BH, and $r$ is the distance between the BBH and the supermassive BH. The gravitational redshift is calculated assuming a non-spinning supermassive BH, although supermassive BHs in AGN disks are typically expected to spin~\citep{2008MNRAS.385.1621K}. This assumption holds, because the effect of spin becomes negligible at the distances relevant for BBHs with respect to the supermassive BH~\citep{2019ApJ...883L...7L,2019PhRvD..99j3005F,2022PhRvD.106j3040C}.

The second effect, which is related to the motion of the BBH as it orbits the supermassive BH, results in a relativistic redshift,
\begin{equation}\label{rel}
z_{\rm rel} = 
\gamma \left[1+v\,{\rm cos}(\phi)\right] - 1 \,,
\end{equation}
where $\gamma = (1-v^2)^{-1/2}$ is the Lorentz factor, $v$ is the magnitude of the velocity, and $\phi$ is the viewing angle between the velocity and the line of sight in the observer frame. Assuming that the BBH is on a circular orbit around a non-spinning supermassive BH~\citep{2020MNRAS.499.2608F}, its velocity is
\begin{equation}
v = 
\frac{1}{\sqrt{2\,(r/R_{\rm s}-1)}} \,.
\end{equation}
Note that~\citep{2023PhRvD.108l3039M} the degeneracy between the mass of the source and the relativistic redshift remains unresolved because detection is mainly done with the dominant quadrupolar mode~\citep{2023PhRvD.107j3044Y}. However, the ambiguity between the relativistic redshift and other redshifts can be eliminated by detecting higher spherical modes, thus allowing, in principle, for the measurement of the source's velocity~\citep{2008PhRvD..78d4024G,2016PhRvD..93h4031B,2021PhRvD.104l3025T}.

In this case, the detected mass of a redshifted source differs from intrinsic mass in the source frame by a factor equal to the combined effects of all relevant redshifts, including gravitational and relativistic redshifts~\citep{2019MNRAS.485L.141C}:
\begin{equation}
m^{z,{\rm eff}} = 
(1+z_{\rm grav}) (1+z_{\rm rel})\,m^z \,,
\end{equation}
where $m^{z,{\rm eff}}$ represents the observed mass of the source in the detector frame, $m^z = (1+z_{\rm c})\,m$ is the observed mass with the cosmological redshift $z_{\rm c}$, in which $m$ is the intrinsic mass in the source frame.

The effective distance of the source is also influenced by various redshift effects, with the relativistic redshift component, induced by the aberration of GWs, being squared in its impact on the observed distance~\citep{2023PhRvD.107d3027T}.
\begin{equation}
D_L^{\rm eff} = 
(1+z_{\rm grav}) (1+z_{\rm rel})^2 \,D_L \,,
\end{equation}
where $D_L = (1+z_{\rm c})\,D_{\rm com}$ is the luminosity distance of the source with $D_{\rm com}$ the comoving distance between source and observer.

Consequently, the environmental effects of the supermassive BH, such as gravitational and relativistic redshifts, can be accounted for by simply replacing the observed mass $m^z$ with the effective mass $m^{z,{\rm eff}}$, and the luminosity distance $D_L$ with the effective luminosity distance $D_L^{\rm eff}$~\citep{2023PhRvD.108l3039M}.

\subsection{Bayesian criterion}\label{bc}
In order to assess whether it is more likely that the two GW events, GW190514 and GW190521 (as a hierarchical triple merger), as well as the EM transient ZTF19abanrhr, originate from a common source, or whether the three detections are associated by chance, we employ a Bayesian statistical analysis. Given three detections in the GW190514 dataset $d_1$, the GW190521 dataset $d_2$, and the ZTF19abanrhr dataset $s$, the \textit{odds} $\mathcal{O}^1_0$ between the association model $\mathcal{H}_1$ and the coincidence model $\mathcal{H}_0$ is given by
\begin{equation}\label{eqOdds}
\mathcal{O}^1_0(d_1,d_2,s) = 
\frac{p(\mathcal{H}_1|d_1,d_2,s)}{p(\mathcal{H}_0|d_1,d_2,s)} = 
\frac{p(d_1,d_2,s|\mathcal{H}_1)}{p(d_1,d_2,s|\mathcal{H}_0)} \,
\frac{p(\mathcal{H}_1)}{p(\mathcal{H}_0)} \,.
\end{equation}

Here 
\begin{equation}\label{eqBF}
\mathcal{B}^1_0(d_1,d_2,s) = 
\frac{p(d_1,d_2,s|\mathcal{H}_1)}{p(d_1,d_2,s|\mathcal{H}_0)} \,,
\end{equation}
is the \textit{Bayes factor} and $\mathcal{P}^1_0 = p(\mathcal{H}_1)/p(\mathcal{H}_0)$ is the \textit{prior odds}. 
The prior odds for the two models can be expressed as the product of two components: 
(i) the odds that the EM transient is associated with one of the two GW events, estimated to be $\sim1/13$~\citep{2021CQGra..38w5004A}, based on the fact that approximately 13 transients similar to ZTF19abanrhr were detected in the Zwicky Transient Facility alert stream over the relevant observing period~\citep{2020PhRvL.124y1102G};    
(ii) the odds that both mergers in the hierarchical triple system are found in the observed BBHs, estimated to be $\sim1/1000$~\citep{2024ApJ...965...80G,2023MNRAS.523.4113T}, corresponding to the expected fraction of hierarchical triple mergers among detectable events in optimistic scenarios. 
The resulting combined astrophysical prior odds is approximately $1/13{,}000$. For robustness, we adopt a slightly more conservative value of $10^{-5}$ in our analysis, noting that the $1/1000$ estimate represents an optimistic upper limit and the typical value is likely smaller. This choice (approximately eight times smaller) ensures that the preference for the proposed association remains significant even under stricter prior assumptions.

The Bayes factor is the ratio of the evidences for the two hypothetical models. The evidence for a model can be computed via the marginalization over all parameters $\vec \theta$ required by the model $\mathcal{H}_i$:
\begin{equation}
p(d_1,d_2,s|\mathcal{H}_i) = 
\int {\rm d}{\vec \theta} \, p(d_1,d_2,s|\vec \theta,\mathcal{H}_i) \, p(\vec \theta|\mathcal{H}_i) \,.
\end{equation}
The parameters of interest for this work are the sky localization $\Omega$ and luminosity distance $D_L$ of the three sources, and the overlap mass $m^{z}$ in the detector frame, which refers to the final mass of GW190514 and to the second mass of GW190521. One has
\begin{widetext}
\begin{equation}\label{eqab}
p(d_1,d_2,s|\mathcal{H}_i) = 
\int {\rm d}D_L {\rm d}\Omega {\rm d}m^z \, 
p(d_1,d_2,s|D_L,\Omega,m^z,\mathcal{H}_i) \,
p(D_L,\Omega,m^z|\mathcal{H}_i) \,.
\end{equation}
\end{widetext}

The first term in Equation\,(\ref{eqab}), $p(d_1,d_2,s|D_L,\Omega,m^z,\mathcal{H}_i)$ can be written as 
\begin{widetext}
\begin{equation} \label{eqft}
p(d_1,d_2,s|D_L,\Omega,m^z,\mathcal{H}_i) \propto
\frac{p(D_L,\Omega,m^z|d_1,\mathcal{H}_i)}{p(D_L,\Omega,m^z|\mathcal{H}_i)} \,
\frac{p(D_L,\Omega,m^z|d_2,\mathcal{H}_i)}{p(D_L,\Omega,m^z|\mathcal{H}_i)} \,
\frac{p(D_L,\Omega|s,\mathcal{H}_i)}{p(D_L,\Omega|\mathcal{H}_i)} \,,
\end{equation}
\end{widetext}
where $p(D_L,\Omega,m^z|d_i,\mathcal{H}_i)$ is the posterior probability density for the GW dataset $d_i$ and $p(D_L,\Omega,m^z|\mathcal{H}_i) = p(D_L,\Omega|\mathcal{H}_i)\,p(m^z|\mathcal{H}_i) = p(D_L|\mathcal{H}_i)\,p(\Omega|\mathcal{H}_i)\,p(m^z|\mathcal{H}_i)$ is the prior probability distribution used during the parameter estimation run by the LIGO-Virgo-KAGRA (LVK) Collaboration. The parameter estimation prior is uniform in the sky localization and mass, and in the square of the luminosity distance. The possible travel between the mergers is neglected~\citep{2021ApJ...907L..48V}. 
The probability density $p(D_L,\Omega|s,\mathcal{H}_i) = p(D_L|s,\mathcal{H}_i)\,p(\Omega|s,\mathcal{H}_i)$ is obtained from the localization of its host {\AGN}. The conversion in $p(D_L|s,\mathcal{H}_i)$ between luminosity distance and redshift takes into account environmental effects, which are discussed in detail in the corresponding models below.

The second term in Equation\,(\ref{eqab}), $p(D_L,\Omega,m^z|\mathcal{H}_i)$, is the prior conditioned under the hypothetical model $\mathcal{H}_i$. In general, these parameters are a \textit{priori} independent; we can factorize it as
\begin{equation} \label{eqst}
p(D_L,\Omega,m^z|\mathcal{H}_i) = 
p(D_L|\mathcal{H}_i)\,
p(\Omega|\mathcal{H}_i)\,
p(m^z|\mathcal{H}_i) \,,
\end{equation}
where the prior $p(m^z|\mathcal{H}_i)$ differs in the different models. In the association model $\mathcal{H}_1$, $p(m^z|\mathcal{H}_1) = p^{\rm 1g}_{m_{\rm f}}(m^z|\mathcal{H}_1)$ represents the final detector-frame mass distributions of first generation BBH mergers that are not a result of a previous merger in the AGN formation channel. In the coincidence model $\mathcal{H}_0$, $p(m^z|\mathcal{H}_0) = p^{\rm 1g}_{m_{\rm f}}(m^z|\mathcal{H}_0)$ and $p(m^z|\mathcal{H}_0) = p^{\rm hier}_{m_{\rm 2}}(m^z|\mathcal{H}_0)$ refer to the final mass distributions of first-generation BBH mergers and the second mass distributions of hierarchical BBH mergers, respectively, in the detector frame, for formation channels other than AGN disks.

Substituting Equations\,(\ref{eqft})\&(\ref{eqst}) into Equation\,(\ref{eqab})
\begin{widetext}
\begin{equation} \label{priorH1}
\begin{aligned}
p(d_1,d_2,s|\mathcal{H}_1) 
& = \int {\rm d}D_L {\rm d}\Omega {\rm d}m^z \, 
\frac{p(D_L,\Omega,m^z|d_1,\mathcal{H}_1)}{p(D_L|\mathcal{H}_1)\,p(\Omega|\mathcal{H}_1)\,p(m^z|\mathcal{H}_1)} \,
\frac{p(D_L,\Omega,m^z|d_2,\mathcal{H}_1)}{p(D_L|\mathcal{H}_1)\,p(\Omega|\mathcal{H}_1)\,p(m^z|\mathcal{H}_1)} \,\\
&~~~ \times
\frac{p(D_L,\Omega|s,\mathcal{H}_1)}{p(D_L|\mathcal{H}_1)\,p(\Omega|\mathcal{H}_1)} \,
p(D_L|\mathcal{H}_1)\,
p(\Omega|\mathcal{H}_1)\,
p^{\rm 1g}_{m_{\rm f}}(m^z|\mathcal{H}_1) \,\\
& = \int {\rm d}D_L {\rm d}\Omega {\rm d}m^z \, 
\frac{p(D_L,\Omega,m^z|d_1,\mathcal{H}_1) \, p(D_L,\Omega,m^z|d_2,\mathcal{H}_1)p(D_L|s,\mathcal{H}_1)\,p(\Omega|s,\mathcal{H}_1)}{p^2(D_L|\mathcal{H}_1)\,p^2(\Omega|\mathcal{H}_1)\,p^2(m^z|\mathcal{H}_1)} \,
p^{\rm 1g}_{m_{\rm f}}(m^z|\mathcal{H}_1) \,,
\end{aligned}
\end{equation}
\end{widetext}
and
\begin{widetext}
\begin{equation} \label{priorH0}
\begin{aligned}
p(d_1,d_2,s|\mathcal{H}_0) 
& = \int {\rm d}D_L {\rm d}\Omega {\rm d}m^z \, 
\frac{p(D_L,\Omega,m^z|d_1,\mathcal{H}_0)}{p(D_L|\mathcal{H}_0)\,p(\Omega|\mathcal{H}_0)\,p(m^z|\mathcal{H}_0)} \, 
p(D_L|\mathcal{H}_0)\,p(\Omega|\mathcal{H}_0)\,p^{\rm 1g}_{m_{\rm f}}(m^z|\mathcal{H}_0)\,\\
&~~~ \times 
\int {\rm d}D_L {\rm d}\Omega {\rm d}m^z \, 
\frac{p(D_L,\Omega,m^z|d_2,\mathcal{H}_0)}{p(D_L|\mathcal{H}_0)\,p(\Omega|\mathcal{H}_0)\,p(m^z|\mathcal{H}_0)} \, 
p(D_L|\mathcal{H}_0)\,p(\Omega|\mathcal{H}_0)\,p^{\rm hier}_{m_{\rm 2}}(m^z|\mathcal{H}_0)\,\\
&~~~ 
\times \int {\rm d}D_L {\rm d}\Omega \, 
\frac{p(D_L,\Omega|s,\mathcal{H}_0)}{p(D_L|\mathcal{H}_0)\,p(\Omega|\mathcal{H}_0)} \, 
p(D_L|\mathcal{H}_0)\,p(\Omega|\mathcal{H}_0)\,\\
& = \int {\rm d}m^z \, 
\frac{p(m^z|d_1,\mathcal{H}_0)}{p(m^z)} \, 
p^{\rm 1g}_{m_{\rm f}}(m^z|\mathcal{H}_0)\,
\times \int {\rm d}m^z \, 
\frac{p(m^z|d_2,\mathcal{H}_0)}{p(m^z)} \, 
p^{\rm hier}_{m_{\rm 2}}(m^z|\mathcal{H}_0) \,.
\end{aligned}
\end{equation}
\end{widetext}
The second equal sign arises from the fact that the integrals over the spatial localization in Equation\,(\ref{priorH0}) equal unity and are therefore omitted.

For comparison, we also investigate scenarios where only two detections share a common source, instead of all three, by removing irrelevant observation data from Equations\,(\ref{priorH1})\&(\ref{priorH0}). 
Specifically, for three detections ($a$, $b$, and $c$), we define the model $\{a,b,c\}$ as corresponding to $\mathcal{H}_0$, meaning the three detections are unrelated, while the model $\{a\mbox{-}b\mbox{-}c\}$ corresponds to $\mathcal{H}_1$, meaning they share a common origin. The model $\{a\mbox{-}b,c\}$ includes three different combinations, denoted as $\mathcal{H}_2$, $\mathcal{H}_3$, and $\mathcal{H}_4$ (see Table\,\ref{tab1}). In the following sections, we introduce the prescriptions for the distributions $p(D_L|s)$ and $p(m^z|\mathcal{H}_i)$ under the two hypothetical models.

\subsection{Association model}\label{am}
In the association model $\mathcal{H}_1$, the GW events GW190514 and GW190521, as a hierarchical triple merger, and the flare ZTF19abanrhr share a common origin. The localization is fixed to the host {\AGN}, and as a result, the observed mass and distance of the source are altered due to environmental effects. In this case, the GW posterior probability densities used in our work are the detector-frame effective primary mass $m^{z,{\rm eff}}$ and effective luminosity distance $D_L^{\rm eff}$. These two parameters are independently measured in GW detection; however, under the common origin hypothesis, they depend on the additional redshifts introduced by environmental effects.

The luminosity distance of the EM source is expressed as $p(D_L|s,\mathcal{H}_1) \rightarrow p(D_L^{\rm eff}|s,\mathcal{H}_1)$:
\begin{widetext}
\begin{equation} \label{eqrg}
p(D_L^{\rm eff}|s,\mathcal{H}_1) = 
\int {\rm d}D_L {\rm d}r {\rm d}\phi \,
\delta
\left[D_L^{\rm eff}-
(1+z_{\rm grav}(r))\,
(1+z_{\rm rel}(r,\phi))^2\,
D_L\right]\,
p(D_L|s,\mathcal{H}_1)\,
p(r)\,p(\phi)\,,
\end{equation}
\end{widetext}
where 
\begin{equation}
p(D_L|s,\mathcal{H}_1) = \int {\rm d}z\,p(D_L|z,\Omega_{\rm cosm})\,p(z|s,\mathcal{H}_1)
\end{equation}
with $\Omega_{\rm cosm}$ a set of cosmological parameters assuming a Planck18 cosmology for $\Omega_{\rm cosm}$~\citep{2020A&A...641A...6P}, and
\begin{equation}
p(z|s,\mathcal{H}_1) = \delta(z-z_{\rm AGN})
\end{equation}
the redshift $z_{\rm AGN} = 0.438$ of {\AGN}\footnote{\url{https://skyserver.sdss.org/dr12/en/tools/explore/Summary.aspx?id=1237665128546631763}}. The expressions for $z_{\rm grav}(r)$ and $z_{\rm rel}(r,\phi)$ in Equation\,(\ref{eqrg}) are the ones in Equations~(\ref{grav}) and (\ref{rel}), respectively.
Here, we adopt the prior $p(\phi)$ to be uniform in ${\rm cos}(\phi)$. 
The radial distance prior, $p(r)$, between the BBH and the supermassive BH is predicted based on migration traps in AGN disks~\citep{2016ApJ...819L..17B}. We assume that this radial distance is uniformly distributed between 24.5 and 331 Schwarzschild radii~\citep{2016ApJ...819L..17B}. It is important to note that the precise location of migration traps can vary depending on the disk model, and in some cases, there may not be a migration trap at all~\citep{2021MNRAS.505.1324P,2024MNRAS.528L.127W,2024MNRAS.530.2114G}. However, our results are robust and are not significantly affected by these uncertainties~\citep{2023PhRvD.108l3039M}.

In the same way, the prior conditioned under the hypothetical model $\mathcal{H}_1$ reads $p(m^z|\mathcal{H}_1) \rightarrow p(m^{z,{\rm eff}}|\mathcal{H}_1)$
\begin{widetext}
\begin{equation}
p(m^{z,{\rm eff}}|\mathcal{H}_1) = 
\int {\rm d}m{\rm d}r {\rm d}\phi \,
\delta \left[m^{z,{\rm eff}}-
(1+z_{\rm grav}(r))\,
(1+z_{\rm rel}(r,\phi))
(1+z_{\rm c})\,m\right] \,
p(m|\mathcal{H}_1) \,
p(r)\,p(\phi)\,,
\end{equation}
\end{widetext}
where the cosmological redshift $z_c \equiv z_{\rm AGN}$ and $p(m^z|\mathcal{H}_1) = (1+z_{\rm c})\,p(m|\mathcal{H}_1)$. The astrophysical prior $p(m|\mathcal{H}_1)$ represents the mass distribution of BBHs formed in AGN disks, which is presented in Section~\ref{ad}.

\subsection{Coincidence model}\label{cm}
In the coincidence model $\mathcal{H}_0$, the three transients GW190514, GW190521, and ZTF19abanrhr are unrelated and happen to appear within the same localization by chance. In this scenario, there are no environmental effects from AGNs, and, therefore, the only redshift present is the cosmological one. As a result, the prior conditioned under the hypothetical model $\mathcal{H}_0$ is given by

\begin{equation} 
p(m^{z}|\mathcal{H}_0) = 
(1+z_{\rm c})\,p(m|\mathcal{H}_0) \,,
\end{equation}
where $p(m|\mathcal{H}_0)$ refers to the mass distribution of BBHs formed from channels other than AGN disks, discussed below.

\subsection{Astrophysical distributions}\label{ad}
To obtain the astrophysical distributions, we utilize a phenomenological population model introduced in~\citet{2023PhRvD.107f3007L}, which allows for the rapid simulation of BBH mergers parametrized, including hierarchical mergers, within dynamical formation channels. For the first-generation BBH merger distributions, we adopt the results from the analysis of GWTC-3 by the LVK Collaboration~\citep{2023PhRvX..13a1048A}.
In particular, the primary masses of BBH mergers that do not result from a previous merger follow a distribution described by a \textsc{PowerLaw+Peak} model. The parameters of the \textsc{PowerLaw+Peak} model used for the coincidence model $\mathcal{H}_0$ are consistent with the results of GWTC-3. For the association model $\mathcal{H}_1$, the mass power-law index aligns with predictions for a top-heavy population of merging BHs in AGN disks~\citep{2019ApJ...876..122Y}.
The dimensionless spin-magnitude distribution of first-generation BHs is uniformly assigned within a beta distribution. In the model $\mathcal{H}_0$, its parameters are obtained from the GWTC-3, while the parameters determining the distribution's mean and variance are set to $1.5$ and $3.0$ in the model $\mathcal{H}_1$, reflecting the angular momentum exchange between binaries and the disk~\citep{2020MNRAS.494.1203M}. In both models, the spin tilt angles are randomly selected from an isotropic distribution across a sphere.

In order to pair BBHs, the mass ratios of the binaries are drawn from a \textsc{PowerLaw} distribution. Similarly, in the model $\mathcal{H}_0$, we take the mass-ratio power-law parameter from the GWTC-3, setting it to be $\sim 0$ in the model $\mathcal{H}_1$, which represents the expectations of a runaway merger scenario in AGN disks~\citep{2022PhRvD.105f3006L}. 
We assume an \texttt{NG+$\leq$NG} branch for hierarchical mergers (see details in~\citet{2023PhRvD.107f3007L}). We use numerical relativity fits to calculate each merger remnant’s final mass, final spin, and kick velocity~\citep{2012ApJ...758...63B,2016ApJ...825L..19H,2007ApJ...659L...5C}. To ensure that our results are not impacted by the potential parameter values used above, we repeated the analysis with variations of these parameters and found that our results do not strongly depend on them.

\subsection{Models including primary masses}\label{mpm}
Following the considerations from previous literature, we take the primary masses of GW events into account in our hypothetical models~\citep{2023PhRvD.108l3039M}.  The Equations~(\ref{priorH1}) and (\ref{priorH0}) become
\begin{widetext}
\begin{equation}
\begin{aligned}
p(d_1,d_2,s|\mathcal{H}_1) = 
& \int {\rm d}D_L {\rm d}\Omega {\rm d}m^z {\rm d}m'^{,z} {\rm d}m''^{,z} \,\\
& \times \frac{p(D_L,\Omega,m^z,m'^{,z}|d_1,\mathcal{H}_1)\,
p(D_L,\Omega,m^z,m''^{,z}|d_2,\mathcal{H}_1)\,
p(D_L|s)\,p(\Omega|s,\mathcal{H}_1)}{p^2(D_L|\mathcal{H}_1)\,p^2(\Omega|\mathcal{H}_1)\,p^2(m^z|\mathcal{H}_1)\,
p(m'^{,z}|\mathcal{H}_1)\,p(m''^{,z}|\mathcal{H}_1)} \,\\
& \times p^{\rm 1g}_{m_{\rm f}}(m^z|\mathcal{H}_1) \, 
p^{\rm 1g}_{m_{\rm 1}}(m'^{,z}|\mathcal{H}_1) \, 
p^{\rm hier}_{m_{\rm 1}}(m''^{,z}|\mathcal{H}_1) \,,
\end{aligned}
\end{equation}
\end{widetext}
and
\begin{widetext}
\begin{equation}
\begin{aligned}
p(d_1,d_2,s|\mathcal{H}_0) = &
\int {\rm d}m^z {\rm d}m'^{,z} \, 
\frac{p(m^z,m'^{,z}|d_1,\mathcal{H}_0)}{p(m^z|\mathcal{H}_0)\,p(m'^{,z}|\mathcal{H}_0)} \, 
p^{\rm 1g}_{m_{\rm f}}(m^z|\mathcal{H}_0) \, p^{\rm 1g}_{m_{\rm 1}}(m'^{,z}|\mathcal{H}_0)\,\\
& \times \int {\rm d}m^z {\rm d}m''^{,z}\, 
\frac{p(m^z,m''^{,z}|d_2,\mathcal{H}_0)}{p(m^z|\mathcal{H}_0)\,p(m''^{,z}|\mathcal{H}_0)} \, 
p^{\rm hier}_{m_{\rm 2}}(m^z|\mathcal{H}_0) \, p^{\rm hier}_{m_{\rm 1}}(m''^{,z}|\mathcal{H}_0) \,,
\end{aligned}
\end{equation}
\end{widetext}
respectively, where $m'^{,z}$ and $m''^{,z}$ are the primary masses of GW190514 and GW190521 in the detector frame, respectively; $p(m'^{,z}|\mathcal{H}_1)$ and $p(m''^{,z}|\mathcal{H}_1)$ are the priors used during the parameter estimation run; and 
$p^{\rm 1g}_{m_{\rm 1}}(m'^{,z}|\mathcal{H}_1)$ and 
$p^{\rm hier}_{m_{\rm 1}}(m''^{,z}|\mathcal{H}_1)$ 
($p^{\rm 1g}_{m_{\rm 1}}(m'^{,z}|\mathcal{H}_0)$ and 
$p^{\rm hier}_{m_{\rm 1}}(m''^{,z}|\mathcal{H}_0)$)
are the priors conditioned of primary detector-frame masses of first generation and hierarchical mergers in the AGN formation channel (in formation channels other than AGN disks), respectively.

\section{Results}\label{sec:result}

\begin{figure}
\centering
\includegraphics[width=8cm]{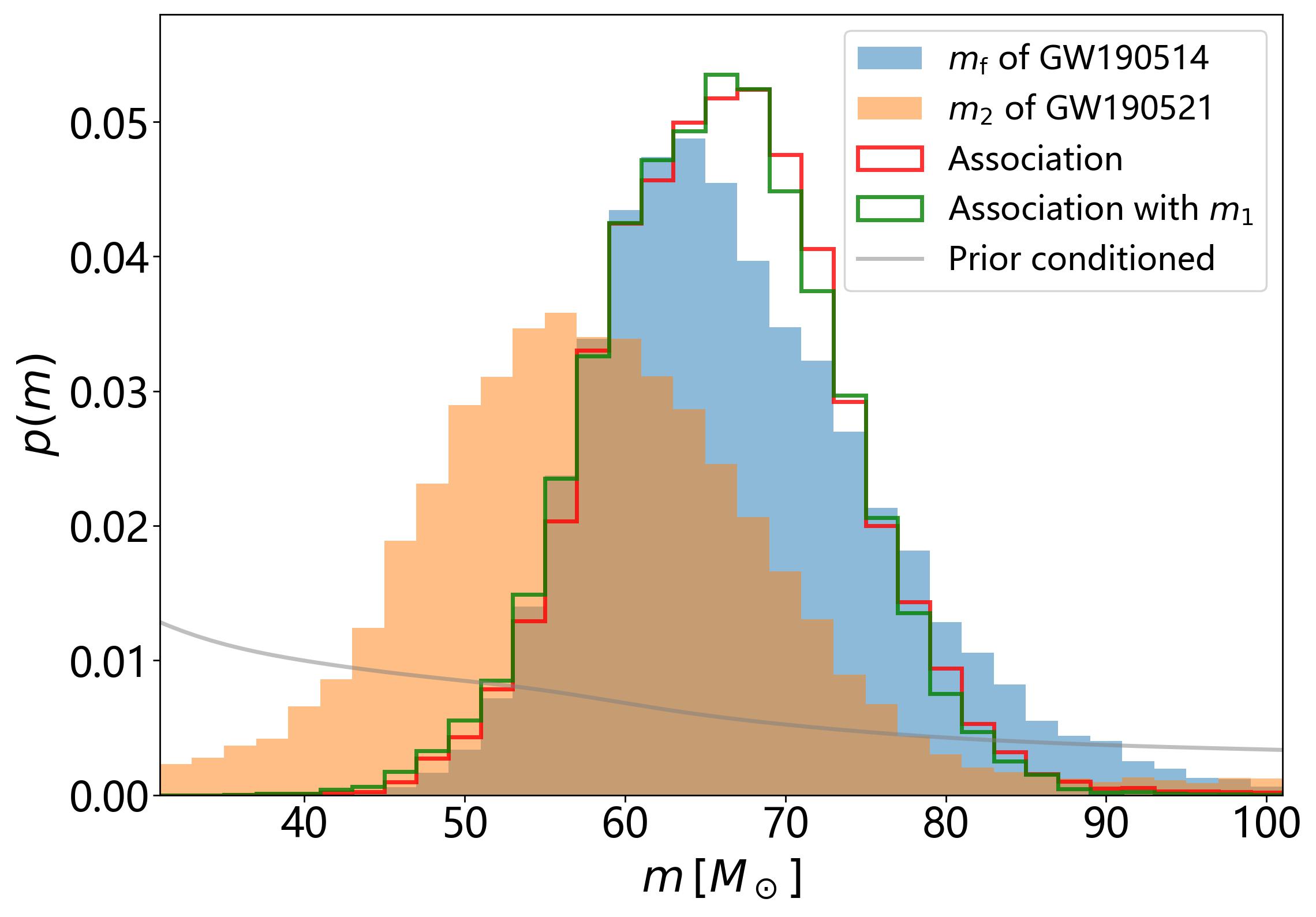}
\caption{
Probability distribution of the associated mass.
The empty histograms, each normalized, illustrate the associated mass distribution of GW190514-GW190521 for the association model, both without (red) and with (green) the inclusion of the primary mass $m_1$. In contrast, the filled histograms represent the final mass distribution of GW190514 (blue) and the second mass distribution of GW190521 (yellow), as obtained from the GWTC-2.1. The gray line indicates the mass prior conditioned for the association model, which corresponds to the final mass distribution of the first generation BBH mergers in the AGN formation channel.
}
\label{fig2} 
\end{figure}

Figure\,\ref{fig2} shows the posterior probability densities for the final mass of GW190514 and the second mass of GW190521, alongside the association model (labeled as `Association', red). The distribution for the association model is more tightly constrained with respect to the coincidence model (as the GWTC-2.1 distribution~\citep{2024PhRvD.109b2001A}). Making use of the standard Bayesian Monte Carlo sampling algorithms, we perform model selection by calculating the evidence for each hypothesis. Accounting for the environmental effects, the log Bayes factor is ${\rm log}\mathcal{B}=16.8$, corresponding to an odds ratio of $\mathcal{O}\sim$$199:1$, assuming an astrophysical prior odds of $\mathcal{P}\sim10^{-5}$, which strongly supports the common origin hypothesis. Through specific analysis, we find that the Bayes factor is driven by two major contributions: i) the consistent localizations, and ii) the different mass priors for the two models. In particular, the log Bayes factors is ${\rm log}\mathcal{B}=8.7$ when considering localizations alone. In addition, to ensure that our results are not biased by differing astrophysical priors or by the absence of the mass gap (caused by (pulsational) pair-instability supernovae~\citep{2003ApJ...591..288H,2007Natur.450..390W}) which are less likely in AGN disks due to increased probabilities for hierarchical mergers and accretion~\citep{2023PhRvD.108l3039M}, we repeated the analysis using a mass model without the mass gap as the prior for the coincidence model, yielding a similar result.

\begin{figure}
\centering
\includegraphics[width=8cm]{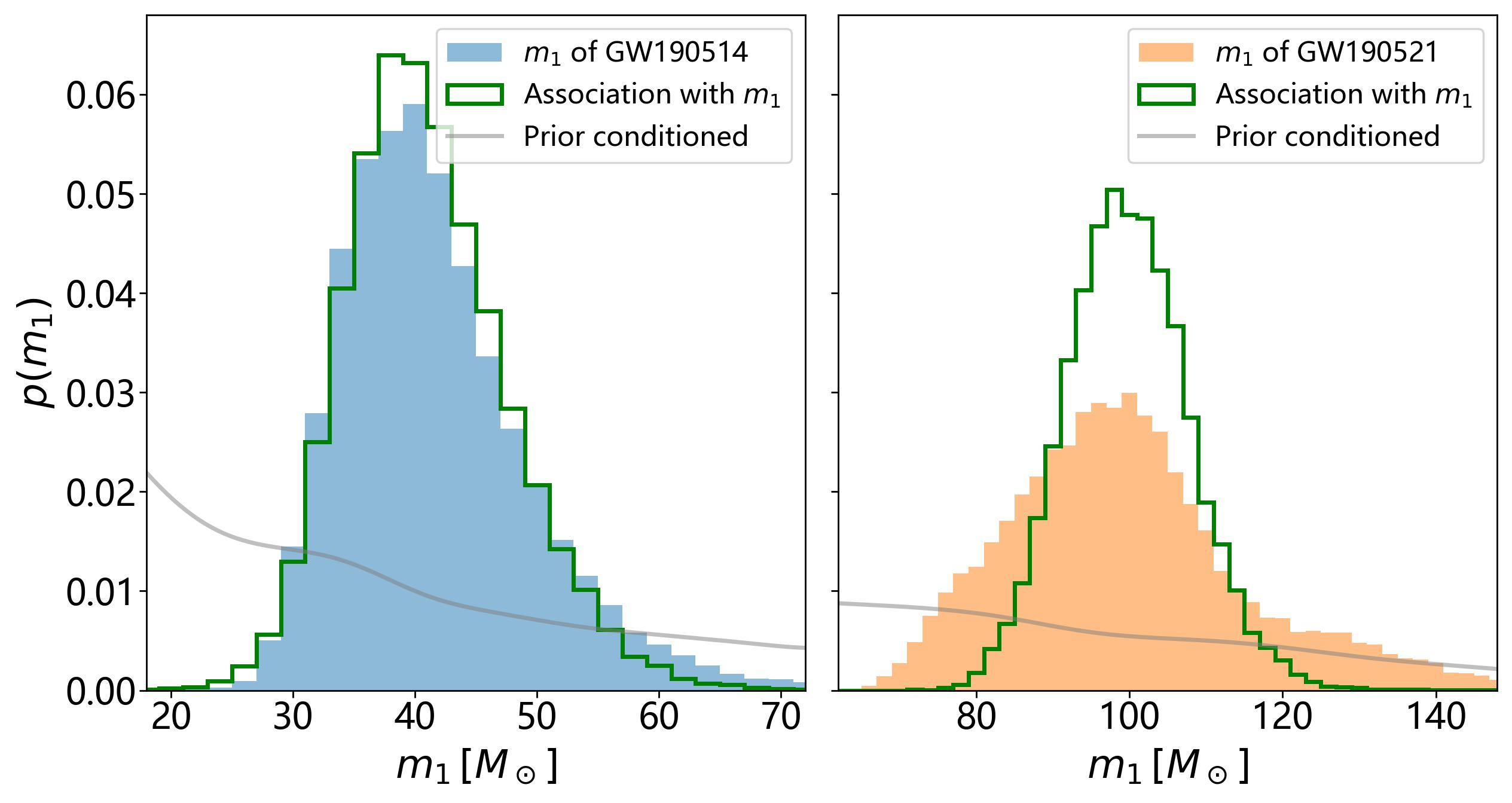}
\caption{
Probability distribution of the primary mass.
The empty histograms, each normalized, represent the primary mass distribution for the association model (green). For comparison, the filled histograms depict the primary mass distribution from the GWTC-2.1. The gray line indicates the mass prior conditioned for the association model. On the left, the distributions for GW190514 are displayed, with the prior conditioned corresponding to the primary mass distribution of first generation BBH mergers in the AGN formation channel. On the right, the distributions for GW190521 are shown, with the prior conditioned reflecting the primary mass distribution of hierarchical BBH mergers in the AGN formation channel.
}
\label{fig3} 
\end{figure}
Furthermore, we find that the primary masses of these two GW events are also crucial for enhancing the probability of the association compared to the coincidence~\citep{2023PhRvD.108l3039M}. This is because BBH mergers in AGN disks are generally expected to have more massive components compared to other channels, such as isolated binary evolution or star clusters, due to accretion and frequent hierarchical mergers~\citep{2019PhRvL.123r1101Y,2022PhRvD.105f3006L,2023PhRvD.107f3007L}. Therefore, we also consider the primary masses ($m_1$) of the GW events to assess the multimessenger coincidence significance for the three transients (see Section~\ref{mpm}).
Figure\,\ref{fig3} presents the posterior probability densities of the primary masses of the two GW events, comparing the distributions from the GWTC-2.1~\citep{2024PhRvD.109b2001A} with those from the association model (labeled as `Association with $m_1$', green). As shown in this figure, the association model with $m_1$ provides better constraints, slightly favoring a more massive BH for GW190521 and a less massive one for GW190514, compared to the GWTC-2.1 distributions~\citep{2024PhRvD.109b2001A}. 
Notably, GW190521's primary BH mass falls within the mass gap with nearly 100\% probability, strongly suggesting that it is a hierarchical merger~\citep{2020ApJ...902L..26F,2020ApJ...900L..13A,2021MNRAS.502.2049L,2021ApJ...915L..35K,2023NatAs...7...11G,2023PhRvD.108h4044A,2024ApJ...975..117M,2024ApJ...977..220A}, 
potentially a second (or higher) -generation merger resulting from GW190514 in an AGN environment. 
Incorporating the primary mass information significantly strengthens the association hypothesis, increasing the Bayes factor by approximately 2 orders of magnitude, corresponding to a log Bayes factor of ${\rm log}\mathcal{B}=21.5$ and an odds ratio of $\mathcal{O}\sim$$2\times10^4:1$. This result aligns with the expectation that more massive BBH mergers are more likely to occur in the AGN formation channel~\citep{2019PhRvL.123r1101Y,2022PhRvD.105f3006L,2023PhRvD.107f3007L}. Additionally, we find that the inclusion of primary mass data allows slightly for better constraints on the associated mass between GW190514 and GW190521 (labeled as `Association with $m_1$', green, in Fig.\,\ref{fig2}).

\begin{table*}
\centering
\caption{\label{tab1}%
Model comparison results.
The model $\mathcal{H}_0$: $\{a,b,c\}$ refers to the scenario where the three detections are unrelated.
The model $\mathcal{H}_1$: $\{a\mbox{-}b\mbox{-}c\}$ corresponds to the case where  the three detections share a common origin.
In contrast, the models $\mathcal{H}_2$, $\mathcal{H}_3$, and $\mathcal{H}_4$: $\{a\mbox{-}b,c\}$ mean that two of the detections (i.e., $a$ and $b$) share a common source. The values in parentheses correspond to hypothetical models with the primary mass information in the AGN formation channel. The log Bayes factors are relative to the model $\mathcal{H}_0$.
}
\begin{tabular}{lccc}\hline
 Models &  Prior-odds & Log-Bayes-factors & Odds\\\hline
$\mathcal{H}_0$: GW190514, GW190521, ZTF19abanrhr & - & 0 & - \\
$\mathcal{H}_1$: GW190514-GW190521-ZTF19abanrhr & $10^{-5}$ & 16.8(21.5) & 199(21772)\\
$\mathcal{H}_2$: GW190514-GW190521, ZTF19abanrhr & 1/1000 & 5.9 & 0.4 \\
$\mathcal{H}_3$: GW190514-ZTF19abanrhr, GW190521 & 1/13 & 3.6(4.1) & 3(5)\\
$\mathcal{H}_4$: GW190521-ZTF19abanrhr, GW190514 & 1/13 & 4.9(7.8) & 11(180)\\\hline
\end{tabular}
\end{table*}
Finally, we note that in addition to the three detection associations, there may also be scenarios where only two detections share a common source. For comparison, we also investigate the scenarios, which has three different combinations, corresponding to the hypotheses $\mathcal{H}_2$, $\mathcal{H}_3$, and $\mathcal{H}_4$ in Table\,\ref{tab1}. 
In particular, we analyze the hierarchical triple merger pair GW190514-GW190521 with ($\mathcal{H}_1$) and without ($\mathcal{H}_2$) the EM counterpart ZTF19abanrhr. Because this candidate itself is not a high confidence counterpart~\citep{2020PhRvL.124y1102G,2023PhRvD.108l3039M,2021arXiv211212481C,2021ApJ...914L..34P,2021CQGra..38w5004A}.
Table\,\ref{tab1} presents the Bayes factors and odds for various association combinations of GW190514, GW190521, and ZTF19abanrhr, indicating that a common origin shared by the three transients is strongly favored based on the observed data sets, in contrast to the associations involving only two of them. 
In particular, we find that compared with random coincidence ($\mathcal{H}_0$), the odds of GW190514 and GW190521 as a hierarchical triple merger without EM counterpart ($\mathcal{H}_2$) is $0.4:1$ (while that with EM counterpart ($\mathcal{H}_1$) is $199:1$).
This suggests that the EM counterpart provides critical discriminatory power, lifting the likelihood of the hierarchical merger hypothesis by more than two orders of magnitude over the null hypothesis, implying that the EM counterpart is pivotal to breaking degeneracies between a true astrophysical signal and chance alignment.
The odds of $\mathcal{H}_1$ is $\sim$14 ($\sim$117) times greater than the odds of $\mathcal{H}_2$, $\mathcal{H}_3$, and $\mathcal{H}_4$ (considering the primary masses).
The analysis results, which account for the localizations and masses of the three transients, provide significantly robust support for the hypothesis of a common origin ($\mathcal{H}_1$).
Moreover, we reanalyze the potential association by including the spins of the GW events. However, we find that the spins do not contribute significant information, yielding a Bayes factor of approximately 1.03 for the spin-only model.

\section{Conclusions and discussion}\label{sec:cd}
In this work, we investigate the potential association between GW190514 and GW190521 as a hierarchical triple merger associated with ZTF19abanrhr, taking into account the sky position, distance, and mass of the sources using a Bayesian criterion. Although the previous study has shown that GW190514 and GW190521 are a plausible hierarchical triple merger, their analysis is not significant with $p$-value of $\sim$0.14~\citep{2021ApJ...907L..48V} (generally, a $p$-value of $\leq$0.05 is considered `significant'). Here, we introduce a new approach and propose using multimessenger~\citep{2020PhRvL.124y1102G} signals to analyze the probability of the GW190514-GW190521 hierarchical triple merger. Our analysis (considering the primary masses) results greatly increase the probability, finding that a log Bayes factor of ${\rm log}\mathcal{B}=16.8~(21.5)$, corresponding to an odds ratio of $\mathcal{O}\sim$$199~(2\times10^4):1$, assuming an astrophysical prior odds of $\mathcal{P}\sim10^{-5}$. This suggest strong evidence for the first hierarchical triple merger associated with an EM counterpart in the AGN formation channel. The finding of the multimessenger hierarchical triple merger in the AGN is crucial for understanding the formation mechanisms of massive BHs, the role of AGNs in hierarchical mergers, and the implications of multimessenger astronomy.

We show that the GW190514-GW190521 pair, which is separated by only one week, further supports the scenario of a triple hierarchical merger chain if associated with ZTF19abanrhr: i) Such a short time frame between the two mergers presents significant challenges for the reformation of a binary system and its subsequent merger. However, AGNs provide an environment conducive to overcoming these challenges~\citep{2017ApJ...835..165B,2018ApJ...866...66M}. Specifically, the number density of BHs in the migration traps~\citep{2016ApJ...819L..17B} of AGN disks is notably high due to migration~\citep{2012MNRAS.425..460M}. This high density facilitates the rapid reformation of a binary system from GW190514's remnant BHs, enabling a swift second merger with the aid of the dense gas in AGN disks;
ii) GW190521 may not have had sufficient time to circularize in such a short interval, suggesting that it is likely to be eccentric~\citep{2022NatAs...6..344G,2020ApJ...903L...5R}. This inference is supported by previous analyses indicating a non-zero eccentricity for GW190521~\citep{2021ApJ...907L..48V}, and AGN disk environments are known to promote an excess of eccentric mergers~\citep{2022Natur.603..237S}. 
Our current methodological framework explicitly excludes time delay and eccentricity considerations in the statistical analysis, as these require dedicated waveform development that will be addressed in subsequent studies. Furthermore, we anticipate environmental effects to induce luminosity distance corrections of $\lesssim 1\%$---a systematic uncertainty. This is tiny compared to the uncertainty in intrinsic luminosity distance for each BBH observation (typically $\pm30\%$).
The impact warrants future investigation through targeted parameter estimation studies~\citep{2025arXiv250318333L}.

Our Bayesian analysis conditions on the association with ZTF19abanrhr, while we observe that multiple candidate EM counterparts to BBH mergers have been reported in the literature~\citep{2023ApJ...942...99G} since the first one for GW190521. This may introduce a potential trials factor that could affect the final significance of single association. However, our analysis focuses on evaluating the hierarchical merger scenario under the assumption that ZTF19abanrhr is a genuine EM counterpart. 
Incorporating a full population-level analysis to account for multiple GW-EM candidates is beyond the scope of this work but would be a valuable extension for future studies. 
The full analysis could include additional astrophysical properties of the AGN (beyond the three-dimensional spatial position of EM counterpart), such as the structure of the gas disk, the mass of the flare, etc, which may further confidently determine the validity of this association.
Nevertheless, the high Bayes factor and odds ratio obtained here provide strong evidence for the plausibility of this particular hierarchical triple merger scenario, highlighting the importance of joint GW-EM analyses in probing BH formation channels.

We use the GW datasets released with GWTC-2.1~\citep{2024PhRvD.109b2001A}, labeled as ``cosm'' and ``IMRPhenomXPHM.'' 
This analysis was made possible following software packages:
NumPy~\citep{harris2020array}, 
SciPy~\citep{2020SciPy-NMeth}, 
Matplotlib~\citep{2007CSE.....9...90H}, 
emcee~\citep{2013PASP..125..306F}, 
IPython~\citep{2007CSE.....9c..21P},
corner~\citep{2016JOSS....1...24F},
seaborn~\citep{Waskom2021},
and Astropy~\citep{2022ApJ...935..167A}.

\section{Acknowledgments} 
We acknowledge valuable input from our anonymous referee which improved the original manuscript. This work is supported by National Key R$\&$D Program of China (2020YFC2201400),  the National Natural Science Foundation of China (grant No. 11922303), and the Fundamental Research Funds for the Central Universities, Wuhan University (grant No. 2042022kf1182). 
This research has made use of data or software obtained from the Gravitational Wave Open Science Center (\url{https://gwosc.org}), a service of the LIGO Scientific Collaboration, the Virgo Collaboration, and KAGRA. This material is based upon work supported by NSF's LIGO Laboratory which is a major facility fully funded by the National Science Foundation, as well as the Science and Technology Facilities Council (STFC) of the United Kingdom, the Max-Planck-Society (MPS), and the State of Niedersachsen/Germany for support of the construction of Advanced LIGO and construction and operation of the GEO600 detector. Additional support for Advanced LIGO was provided by the Australian Research Council. Virgo is funded, through the European Gravitational Observatory (EGO), by the French Centre National de Recherche Scientifique (CNRS), the Italian Istituto Nazionale di Fisica Nucleare (INFN) and the Dutch Nikhef, with contributions by institutions from Belgium, Germany, Greece, Hungary, Ireland, Japan, Monaco, Poland, Portugal, Spain. KAGRA is supported by Ministry of Education, Culture, Sports, Science and Technology (MEXT), Japan Society for the Promotion of Science (JSPS) in Japan; National Research Foundation (NRF) and Ministry of Science and ICT (MSIT) in Korea; Academia Sinica (AS) and National Science and Technology Council (NSTC) in Taiwan.

\newcommand{\jcap}{J. Cosmol. Astropart. Phys.}
\newcommand{\physrep}{Phys. Rep.}
\newcommand{\mnras}{Mon. Not. R. Astron. Soc.}
\newcommand{\araa}{Annu. Rev. Astron. Astrophys.}
\newcommand{\aap}{Astron. Astrophys}
\newcommand{\aj}{Astron. J.}
\newcommand{\plb}{Phys. Lett. B}
\newcommand{\apjs}{Astrophys. J. Suppl.}
\newcommand{\app} {Astropart. Phys.}
\newcommand{\apjl}{Astrophys. J. Lett}
\newcommand{\pasp}{Publ. Astron. Soc. Pac.}

%

\end{document}